# An Alternating Direction Algorithm for Hybrid Precoding and Combining in Millimeter Wave MIMO Systems

Nuno Souto,[1,2], João Silva[1,2], João Pavia,[1,2] and Marco Ribeiro,[1,2]

[1] Department of Information Science and Technology, ISCTE-University Institute of Lisbon, 1649-026 Lisboa, Portugal.
[2] Instituto de Telecomunicações, 1049 - 001 Lisboa, Portugal.

Correspondence should be addressed to Nuno Souto: nuno.souto@lx.it.pt

## Abstract

Millimeter-wave (mmWave) technology is one of the most promising candidates for future wireless communication systems as it can offer large underutilized bandwidths and eases the implementation of large antenna arrays which are required to help overcome the severe signal attenuation that occurs at these frequencies. To reduce the high cost and power consumption of a fully digital mmWave precoder and combiner, hybrid analog/digital designs based on analog phase shifters are often adopted. In this work we derive an iterative algorithm for the hybrid precoding and combining design for spatial multiplexing in mmWave massive multiple-input multiple-output (MIMO) systems. To cope with the difficulty of handling the hardware constraint imposed by the analog phase shifters we use the alternating direction method of the multipliers (ADMM) to split the hybrid design problem into a sequence of smaller subproblems. This results in an iterative algorithm where the design of the analog precoder/combiner consists of a closed form solution followed by a simple projection over the set of matrices with equal magnitude elements. It is initially developed for the fully-connected structure and then extended to the partially-connected architecture which allows simpler hardware implementation. Furthermore, to cope with the more likely wideband scenarios where the channel is frequency selective, we also extend the algorithm to an orthogonal frequency division multiplexing (OFDM) based mmWave system. Simulation results in different scenarios show that the proposed design algorithms are capable of achieving performances close to the optimal fully digital solution and can work with a broad range of configuration of antennas, RF chains and data streams.

## I. INTRODUCTION

In order to cope with the ever-increasing demand for higher data rates, future wireless networks must exploit novel wireless technologies. Two of the most promising key technologies for fifth generation (5G) networks are massive multiple-input multiple-output (MIMO) schemes and millimeter-wave communications (mmWave) [1]-[3], which can fulfill the requirements of a wide range of different scenarios, from ultra-dense networks [4] to smart rail mobility [5] . While massive MIMO can improve the spectral efficiency, mmWave bands, ranging from 30 GHz to 300 GHz, enable the access to large underutilized bandwidths for wireless transmissions and at the same time facilitate the implementation of compact large antenna arrays due to the very small wavelength. One of the main challenges regarding mmWave communications lies on the huge path and penetration losses that occur on these frequencies [2]. Still, the ease of implementation of large antenna arrays combined with the application of mmWave MIMO precoding techniques can provide highly directional beams, helping to overcome the severe signal attenuation that occurs at these frequencies [6].

While fully digital precoders and combiners are currently unfeasible due to the need for a dedicated radio frequency (RF) chain per antenna with costly and power hungry mmWave mixed-signal components, the alternative analog beamforming based on simpler analog phase shifters limits the transmission to a single stream. As a compromise between both approaches, a hybrid analog/digital design was proposed in [7]. Regarding the hybrid design, it was shown in [8] that the number of RF chains only needs to be twice the number of data streams in order to achieve the exact performance of the fully digital beamforming architecture. To deal with the cases where the number of RF chains is lower than twice the number of data streams, several heuristic algorithms have been proposed covering different scenarios namely, point-to-point MIMO [7],[9]-[15], multiuser downlink MIMO with single [16] or multiple receiver antennas per user [8],[17]-[19], multiuser uplink [20], multi-cell multiuser [21], general multiuser MIMO interference channels [22] and relay-assisted mmWave systems [23][24]. In this paper we will focus our study on point-to-point MIMO systems. Regarding this system model, in [7] the authors exploited the sparse nature of mmWave channels to formulate the design problem as a sparse signal reconstruction problem which could then be solved using orthogonal matching pursuit (OMP) based algorithms [9]. This approach was extended to the case of imperfect channel knowledge in [10], followed by several alternative heuristic algorithms [11]-[15]. In [11], an algorithm was proposed for designing transceiver hybrid beamformers for rate maximization when the number of data streams and RF chains is the same. The approach in [12] relied on the sequential update of the phases in the RF precoder in a greedy manner in order to solve a weighted nonlinear least-squares problem formulation. In [13], four design methods were presented which can provide different trade-offs between computational complexity and performance. However, the best solution can only be used when the number of streams and RF chains is the same. An extension of the scheme employed in [7] was given in [14], where the output of the OMP algorithm is used for updating the RF precoder which is then applied as an input to the OMP algorithm. In [15] the hybrid precoders design was formulated as a block-sparse reconstruction problem and a low complexity algorithm based on the greedy sequence clustering was proposed for finding the solution.

Most of these hybrid design solutions assume a fully-connected structure where each RF chain is connected to all the antennas through different phase shifters. Unfortunately, this approach can render the implementation complexity very high. By sacrificing some beamforming gain, a simpler hardware implementation can be possible if a partially-connected architecture is adopted [25]-[27]. In this case, each RF chain is connected through phase shifters to only a dedicated subset of the antenna array. In [25], a near-optimal iterative hybrid precoding scheme was presented for the partially-connected structure which relied on the idea of successive interference cancelation (SIC). The digital precoder only allocates power to the different data streams, constraining the number of streams and RF chains to be the same. In [26], the authors present a hierarchical approach where the analog precoder is first determined and then the digital precoder is obtained using a water-filling algorithm. For the analog precoder design, [26] presents two alternative schemes, depending on high or low signal to noise ratio conditions. Using an alternating minimization approach, the authors in [27] propose optimal solutions to both subproblems of analog and digital precoder design. Due to the fact that the digital precoder computation relies on the solution of a semidefinite relaxation problem, the computational complexity can grow very fast with the problem size.

Most of the works described previously focused on narrowband channel. However, due to the large bandwidth available in mmWave bands, it is likely that practical MIMO systems will have to operate in frequency selective channels. Therefore it is important to devise hybrid schemes for wideband mmWave systems. To cope with the multipath fading in this type of channels, multicarrier schemes like orthogonal frequency division multiplexing (OFDM) are often adopted [23][27]. Within the context of relay-aided communications, the authors in [23] applied the sparse approximation framework from [7] and proposed a OMP-based hybrid precoder/combiner for OFDM based mmWave systems. In [27], three different algorithms were developed for OFDM systems based on the principle of alternating minimization: one based on manifold optimization, a lower complexity one based on phase extraction (PE-AltMin) and a third one based on a semidefinite relaxation problem. Even though the three algorithms can approach the performance of the fully digital precoder, the first and third algorithms can incur in substantial computational complexity.

In this work, we adopt a different approach for obtaining a near-optimal solution that can be applied to any configuration of antennas, RF chains and data streams. Starting from the hybrid precoder/combiner design for point-to-point MIMO systems formulated as a matrix factorization problem with unit modulus constraints, as in [7], we derive an iterative algorithm using the alternating direction method of the multipliers (ADMM) as a heuristic for providing fast and good quality solutions. ADMM is a well-known operator splitting method, often adopted for convex optimization problems [28], but can also be a powerful heuristic for several nonconvex problems [28][29]. In here we apply ADMM in order to split the hybrid precoding design problem into a sequence of smaller subproblems with simpler solutions. The main contributions are summarized as follows:

- Starting from a narrowband MIMO channel and a fully-connected structure we show that addressing the hybrid design as a matrix factorization problem and rewriting it in a convenient form allows us to apply ADMM and obtain a natural splitting between the design of the analog and digital parts. Furthermore, the separate design steps have straightforward close-form solutions.

- Although the hybrid design approach employed for the fully-connected structure can be directly extended to the partially-connected case by simply adapting the analog projection step to the set of matrices matching the partially-connected structure, we can develop a lower complexity version of the algorithm by including the special structure of the analog precoder/combiner directly into the original formulation of the problem.

- To cope with frequency selective channels, we extend the proposed hybrid precoder/combiner design to a OFDM-based mmWave system. The hybrid design is formulated as an extension of the matrix factorization problem used for the narrowband case which allows us to obtain a very similar algorithm with only a small modification in the closed-form solution required in two of the steps of the original algorithm.

- Simulation results show that the proposed approach can attain spectral efficiencies close to the optimal fully digital design with a small number of RF chains. Furthermore, it can achieve a better performance-complexity trade-off than other existing methods, in particular when the number of streams and RF chains are different.

The remainder of the paper is organized as follows: section II introduces the model for a narrowband mmWave system and formulates the hybrid precoding and combining design as a matrix factorization problem. Section III derives the proposed algorithm assuming a fully connected structure and evaluates its computational complexity. The hybrid design algorithm is extended to a partially connected architecture in section V and to a OFDM-based mmWave system operating in a frequency selective channel in section VI. Performance results obtained with the proposed algorithms are presented in section IV followed by the conclusions in section VII.

*Notation:* Matrices and vectors are denoted by uppercase and lowercase boldface letters, respectively. The superscript $(\cdot)^H$ denotes the conjugate transpose of a matrix/vector, $\|\cdot\|_F$ is the Frobenius norm, $|\cdot|$ is the determinant, $\text{tr}(\cdot)$ is the trace of a matrix, $\text{blkdiag}\{\ \}$ represents a block diagonal matrix whose elements are the vectors contained in the argument and $\mathbf{I}_n$ is the $n\times n$ identity matrix. $X_{m,n}$ denotes the element on row $m$, column $n$ of matrix $\mathbf{X}$, $\mathbf{X}_{i,:}$ is its $i^{th}$ row and $\mathbf{X}_{:,j}$ is its $j^{th}$ column.

## II. System Model and Problem Statement

Let us consider a mmWave hybrid single-user MIMO system similar to [7] and shown in Fig. 1, where $N_{tx}$ antennas transmit $N_s$ data streams to a receiver with $N_{rx}$ antennas. The number of RF chains, $N_{RF}$, is assumed to be the same at the transmitter and receiver (to simplify the notation) and satisfies $N_s \leq N_{RF} \leq \min(N_{tx}, N_{rx})$. The $N_s \times 1$ symbol vector $\mathbf{s}$ (with $\mathrm{E}\left[\mathbf{s}\mathbf{s}^H\right] = 1/N_s \cdot \mathbf{I}_{N_s}$,) is first precoded by an $N_{RF} \times N_s$ baseband precoding matrix $\mathbf{F}_{BB}$, followed by an RF precoding step with analog phase shifters represented using an $N_{tx} \times N_{RF}$ matrix $\mathbf{F}_{RF}$. Assuming the fully-connected structure of Fig. 1, where each RF chain is connected to all the antennas, then all the elements in $\mathbf{F}_{RF}$ have equal magnitude. The receiver follows a mirrored approach where the received signal is processed using an $N_{rx} \times N_{RF}$ RF analog combining matrix $\mathbf{W}_{RF}$ followed by an $N_{RF} \times N_s$ baseband combining matrix $\mathbf{W}_{BB}$. The resulting signal, $\mathbf{y} \in \mathbb{C}^{N_s \times 1}$, can be written as

$$\mathbf{y} = \sqrt{\varepsilon}\mathbf{W}_{BB}^H\mathbf{W}_{RF}^H\mathbf{H}\mathbf{F}_{RF}\mathbf{F}_{BB}\mathbf{s} + \mathbf{W}_{BB}^H\mathbf{W}_{RF}^H\mathbf{n}, \quad (1)$$

where $\varepsilon$ is the average received power, $\mathbf{H} \in \mathbb{C}^{N_{rx} \times N_{tx}}$ is the channel matrix (assumed to be perfectly known at the transmitter and receiver) and $\mathbf{n} \in \mathbb{C}^{N_{rx} \times 1}$ contains independent zero-mean circularly symmetric Gaussian noise samples with covariance $\sigma_n^2 \mathbf{I}_{N_{rx}}$. In the case of Gaussian signaling, the spectral efficiency achieved by the system is [7]

$$R = \log_2\left|\mathbf{I}_{N_s} + \frac{\varepsilon}{N_s}\mathbf{R}_n^{-1}\mathbf{W}_{BB}^H\mathbf{W}_{RF}^H\mathbf{H}\mathbf{F}_{RF}\mathbf{F}_{BB}\mathbf{F}_{BB}^H\mathbf{F}_{RF}^H\mathbf{H}^H\mathbf{W}_{RF}\mathbf{W}_{BB}\right| \quad (2)$$

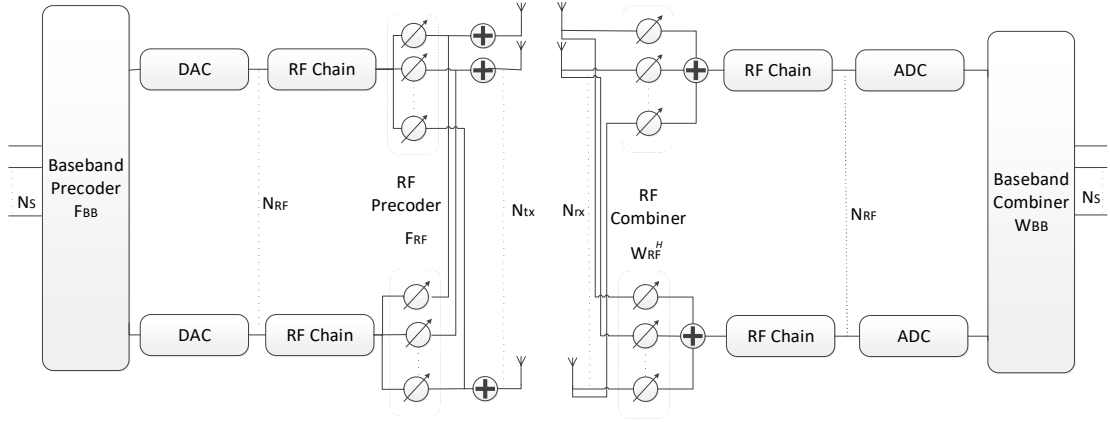

Fig. 1 Block diagram of a mmWave single-user system with hybrid analog/digital precoding and combining. Fully-connected structure.

where $\mathbf{R}_n = \sigma_n^2 \mathbf{W}_{BB}^H \mathbf{W}_{RF}^H \mathbf{W}_{RF} \mathbf{W}_{BB}$. Assuming that the receiver can perform optimal data decoding using the received signal **y**, the maximization of *R* can be well approximated considering the transmitter and receiver sides separately. Let us define $\mathbf{F}_{opt}$ as the unconstrained optimum precoding matrix, formed using the first $N_s$ columns of matrix **V** which is obtained from the singular value decomposition $\mathbf{H} = \mathbf{U\Sigma V}^H$ of the channel ($\mathbf{\Sigma}$ is a diagonal matrix with the singular values in decreasing order). Assuming that the system and channel parameters ($N_{tx}$, $N_{rx}$, $N_{RF}$, number of propagation paths, …) are such that allow the design of $\mathbf{F}_{RF}$ and $\mathbf{F}_{BB}$ satisfying $\mathbf{F}_{opt}^H \mathbf{F}_{RF} \mathbf{F}_{BB} \approx \mathbf{I}_{N_s}$, it was shown in [7] that the hybrid precoding matrices that maximize the data rate are obtained as the solutions of the following nonconvex optimization problem

$$\min_{\mathbf{F}_{RF}, \mathbf{F}_{BB}} f(\mathbf{F}_{RF}, \mathbf{F}_{BB}) \triangleq \left\| \mathbf{F}_{opt} - \mathbf{F}_{RF} \mathbf{F}_{BB} \right\|_F^2 \qquad (3)$$

$$\text{subject to } \mathbf{F}_{RF} \in \mathcal{U}_{N_{tx}, N_{RF}} \qquad (4)$$

$$\left\| \mathbf{F}_{RF} \mathbf{F}_{BB} \right\|_F^2 = N_s \qquad (5)$$

where $\|\cdot\|_F$ is the Frobenius norm, (5) enforces the transmitter's total power constraint and $\mathcal{U}_{N_{tx}, N_{RF}} = \left\{ \mathbf{X} \in \mathbb{C}^{N_{tx} \times N_{RF}} : X_{m,n} = e^{j\theta_{m,n}}, \theta_{m,n} \in [0, 2\pi) \right\}$ is the set of all $N_{tx} \times N_{RF}$ matrices with unit magnitude elements. While, for ease of exposition, we limit the presentation to the hybrid precoder design, the algorithm tries to solve a matrix factorization problem and, therefore, can also be directly applied to the hybrid combiner design. In this case, if the combiner is designed to maximize the data rate, the optimal combiner $\mathbf{W}_{opt}$ consists of the first $N_s$ left singular vectors of $\mathbf{H}$. Alternatively, the combiner can also be designed to minimize the mean squared error between transmitted and received signals [7]. It is important to note that, for the hybrid combiner design, the total power constraint (5) is not required. Although the approach is independent of a specific channel, we adopt the following clustered narrowband model based on the extended Saleh-Valenzuela geometric structure [30][7] with $N_{cl}$ scattering clusters, each with $N_{ray}$ propagations paths:

$$\mathbf{H} = \gamma \sum_{i=1}^{N_{cl}} \sum_{l=1}^{N_{ray}} \alpha_{i,l} \mathbf{a}_r\left(\phi_{i,l}^r, \theta_{i,l}^r\right) \mathbf{a}_t\left(\phi_{i,l}^t, \theta_{i,l}^t\right)^H, \qquad (6)$$

where $\alpha_{i,l}$ is the complex gain of the $l^{th}$ ray from cluster $i$ and $\gamma$ is a normalizing factor such that $\mathrm{E}\left[\|\mathbf{H}\|_F^2\right] = N_{tx} N_{rx}$. Vectors $\mathbf{a}_t\left(\phi_{i,l}^t, \theta_{i,l}^t\right)$ and $\mathbf{a}_r\left(\phi_{i,l}^r, \theta_{i,l}^r\right)$ represent the transmit and receive antenna array responses at the azimuth and elevation angles of $\left(\phi_{i,l}^t, \theta_{i,l}^t\right)$ and $\left(\phi_{i,l}^r, \theta_{i,l}^r\right)$, respectively.

## III. Proposed Hybrid Precoding Algorithm for the Fully-connected Structure

### A. Algorithm Derivation

To assist the application of ADMM to problem (3)-(5) we introduce one auxiliary variable, $\mathbf{R} \in \mathbb{C}^{N_{tx} \times N_{RF}}$, which will allow us to obtain a convenient splitting. Using this variable, we rewrite the problem as

$$\min_{\mathbf{F}_{\mathrm{RF}},\mathbf{F}_{\mathrm{BB}},\mathbf{R},\mathbf{B}} \tilde{f}\left(\mathbf{F}_{\mathrm{RF}},\mathbf{F}_{\mathrm{BB}},\mathbf{R}\right) \triangleq \left\|\mathbf{F}_{opt} - \mathbf{F}_{\mathrm{RF}}\mathbf{F}_{\mathrm{BB}}\right\|_F^2 + I_{\mathcal{U}_{N_{tx},N_{RF}}}(\mathbf{R}) \quad (7)$$

$$\text{subject to} \quad \mathbf{F}_{RF} - \mathbf{R} = 0 \quad (8)$$

$$\left\|\mathbf{F}_{\mathrm{RF}}\mathbf{F}_{\mathrm{BB}}\right\|_F^2 = N_s. \quad (9)$$

where $I_{\mathcal{U}_{N_{tx},N_{RF}}}(\mathbf{R})$ is the indicator function for set $\mathcal{U}_{N_{tx},N_{RF}}$, returning 0 if $\mathbf{R} \in \mathcal{U}_{N_{tx},N_{RF}}$ and $+\infty$ otherwise. The augmented Lagrangian function for (7)-(8) (constraint (9) is handled at the end of the main algorithm similarly to what is done in [7] and [13]) is

$$L_\rho(\mathbf{F}_{\mathrm{RF}},\mathbf{F}_{\mathrm{BB}},\mathbf{R},\mathbf{\Lambda}) = \left\|\mathbf{F}_{opt} - \mathbf{F}_{\mathrm{RF}}\mathbf{F}_{\mathrm{BB}}\right\|_F^2 + I_{\mathcal{U}_{N_{tx},N_{RF}}}(\mathbf{R}) +$$

$$+ 2\operatorname{Re}\left\{\operatorname{tr}\left(\mathbf{\Lambda}^H \left(\mathbf{F}_{RF} - \mathbf{R}\right)\right)\right\} + \rho \left\|\mathbf{F}_{RF} - \mathbf{R}\right\|_F^2, \quad (10)$$

where $\mathbf{\Lambda} \in \mathbb{C}^{N_{tx} \times N_{RF}}$ is the dual variable and $\rho$ is a penalty parameter for constraint (8). For convenience, we can work with a scaled dual variable, $\mathbf{W} = 1/\rho \cdot \mathbf{\Lambda}$, and rewrite the augmented Lagrangian as

$$L_\rho(\mathbf{F}_{\mathrm{RF}},\mathbf{F}_{\mathrm{BB}},\mathbf{R},\mathbf{W}) = \left\|\mathbf{F}_{opt} - \mathbf{F}_{\mathrm{RF}}\mathbf{F}_{\mathrm{BB}}\right\|_F^2 + I_{\mathcal{U}_{N_{tx},N_{RF}}}(\mathbf{R}) + \rho\left\|\mathbf{F}_{RF} - \mathbf{R} + \mathbf{W}\right\|_F^2 - \rho\left\|\mathbf{W}\right\|_F^2. \quad (11)$$

Applying gradient ascent to the dual problem [28] results in the following sequence of iterative steps involving the independent minimization of the augmented Lagrangian over $\mathbf{F}_{\mathrm{RF}}$, $\mathbf{F}_{\mathrm{BB}}$, $\mathbf{R}$ and the update of $\mathbf{W}$. The different steps are detailed next.

• *Step 1: Minimization over* $\mathbf{F}_{\mathrm{RF}}$. Matrix estimate $\mathbf{F}_{\mathrm{RF}}$ for iteration $t+1$ is obtained from

$$\mathbf{F}_{RF}^{(t+1)} = \min_{\mathbf{F}_{RF}} L_\rho(\mathbf{F}_{\mathrm{RF}},\mathbf{F}_{\mathrm{BB}}^{(t)},\mathbf{R}^{(t)},\mathbf{W}^{(t)}). \quad (12)$$

The solution can be found using $\nabla_{\mathbf{F}_{RF}^H} L_\rho(\mathbf{F}_{\mathrm{RF}},\mathbf{F}_{\mathrm{BB}}^{(t)},\mathbf{R}^{(t)},\mathbf{W}^{(t)}) = 0$ from which it is straightforward to show that the following closed form expression is obtained

$$\mathbf{F}_{RF}^{(t+1)} = \left[\mathbf{F}_{opt}\mathbf{F}_{\mathrm{BB}}^{(t)H} + \rho\left(\mathbf{R}^{(t)} - \mathbf{W}^{(t)}\right)\right]\left(\mathbf{F}_{\mathrm{BB}}^{(t)}\mathbf{F}_{\mathrm{BB}}^{(t)H} + \rho\mathbf{I}_{N_{RF}}\right)^{-1}. \quad (13)$$

- *Step 2: Minimization over $\mathbf{F}_{BB}$.* Similarly to step 1, matrix estimate $\mathbf{F}_{BB}$ for iteration $t+1$ is obtained from

$$\mathbf{F}_{BB}^{(t+1)} = \min_{\mathbf{F}_{BB}} L_\rho(\mathbf{F}_{RF}^{(t+1)}, \mathbf{F}_{BB}, \mathbf{R}^{(t)}, \mathbf{W}^{(t)}). \tag{14}$$

Applying $\nabla_{\mathbf{F}_{BB}^H} L_\rho(\mathbf{F}_{RF}^{(t+1)}, \mathbf{F}_{BB}, \mathbf{R}^{(t)}, \mathbf{W}^{(t)}) = 0$ allows us to arrive at the closed form solution

$$\mathbf{F}_{BB}^{(t+1)} = \left(\mathbf{F}_{RF}^{(t+1)H} \mathbf{F}_{RF}^{(t+1)}\right)^{-1} \mathbf{F}_{RF}^{(t+1)H} \mathbf{F}_{opt}. \tag{15}$$

- *Step 3: Minimization over $\mathbf{R}$.* The minimization of (11) with respect to $\mathbf{R}$ can be written as

$$\begin{aligned}\mathbf{R}^{(t+1)} &= \min_{\mathbf{R}} \left\{ I_{\mathcal{U}_{N_{tx}, N_{RF}}}(\mathbf{R}) + \rho \left\| \mathbf{F}_{RF}^{(t+1)} - \mathbf{R} + \mathbf{W}^{(t)} \right\|_F^2 \right\} \\ &= \Pi_{\mathcal{U}_{N_{tx}, N_{RF}}} \left(\mathbf{F}_{RF}^{(t+1)} + \mathbf{W}^{(t)}\right)\end{aligned} \tag{16}$$

where $\Pi_\mathcal{D}(\cdot)$ denotes the projection onto a set $\mathcal{D}$. In the case of set $\mathcal{U}_{N_{tx}, N_{RF}}$, it can be shown that this projection is equivalent to

$$\mathbf{R}^{(t+1)} = \left(\mathbf{F}_{RF}^{(t+1)} + \mathbf{W}^{(t)}\right) \oslash \left|\mathbf{F}_{RF}^{(t+1)} + \mathbf{W}^{(t)}\right|, \tag{17}$$

where $|\cdot|$ is used for denoting elementwise magnitude and $\oslash$ is the Hadamard (or entrywise) division.

- *Step 4: Dual variable update.* The expression for the update of dual variable $\mathbf{W}$ is given by

$$\mathbf{W}^{(t+1)} = \mathbf{W}^{(t)} + \mathbf{F}_{RF}^{(t+1)} - \mathbf{R}^{(t+1)}. \tag{18}$$

Table I summarizes the steps of the proposed algorithm, where $\hat{\mathbf{F}}_{BB}$ and $\hat{\mathbf{F}}_{RF}$ are the final precoding matrices and $Q$ is the maximum number of iterations. Step 10 ensures that constraint (9) is satisfied.

TABLE I

ITERATIVE HYBRID DESIGN ALGORITHM FOR THE FULLY-CONNECTED STRUCTURE

| | |
|---|---|
| 1: | **Input:** $\mathbf{F}_{opt}, \mathbf{F}_{RF}^{(0)}, \mathbf{F}_{BB}^{(0)}, \mathbf{R}^{(0)}, \mathbf{W}^{(0)}, \rho, Q$ |
| 2: | **for** $t=0,1,\ldots Q-1$ **do** |
| 3: | Compute $\mathbf{F}_{RF}^{(t+1)}$ using (13). |
| 4: | Compute $\mathbf{F}_{BB}^{(t+1)}$ using (15). |
| 5: | Compute $\mathbf{R}^{(t+1)}$ via projection (17). |
| 6: | Update $\mathbf{W}^{(t+1)}$ using (18). |
| 7: | **end for**. |
| 8: | $\hat{\mathbf{F}}_{RF} \leftarrow \mathbf{R}^{(Q)}$ |
| 9: | $\hat{\mathbf{F}}_{BB} \leftarrow \left(\hat{\mathbf{F}}_{RF}^{H}\hat{\mathbf{F}}_{RF}\right)^{-1}\hat{\mathbf{F}}_{RF}^{H}\mathbf{F}_{opt}$ |
| 10: | $\hat{\mathbf{F}}_{BB} \leftarrow \sqrt{N_s}\left\|\hat{\mathbf{F}}_{RF}\hat{\mathbf{F}}_{BB}\right\|_F^{-1}\hat{\mathbf{F}}_{BB}$ (for precoder) |
| 11: | **Output:** $\hat{\mathbf{F}}_{BB}, \hat{\mathbf{F}}_{RF}$. |

It is important to note that even though the power constraint (9) is only handled at the end of the main algorithm, it is possible to include it directly inside the iterations. It is only necessary to add an additional auxiliary variable that matches $\mathbf{F}_{RF}\mathbf{F}_{BB}$ and integrate constraint (9) into the objective function (7) using the indicator function for the set of matrices with Frobenius norm equal to $N_s$. Applying ADMM to this formulation results in two additional steps: one minimization of the augmented Lagrangian over the new auxiliary variable, which can be implemented simply as the projection over an hypersphere of radius $\sqrt{N_s}$; the update of a new dual variable (related to the equality constraint involving the new auxiliary variable and $F_{RF}F_{BB}$) which has the same form as (18). It was observed, however, that using the simpler algorithm version presented in Table I results in negligible performance loss. This is also in agreement with lemma 1 from [27], which states that if the Euclidean distance between the non-normalized hybrid precoders and the optimal precoder is sufficiently small, the distance will still be small after the normalization step.

While we do not consider the effect of quantized phase shifters, the proposed algorithm can directly cope with this additional constraint by performing projection (16) over a discretized version of set $\mathcal{U}_{N_{tx},N_{RF}}$, which corresponds to finding the closest element in that set (it can also be approximated through element-wise quantization).

### B. Initialization and Termination

For the initialization, $\mathbf{F}_{RF}^{(0)}$ can be randomly selected from $\mathcal{U}_{N_{tx},N_{RF}}$ followed by $\mathbf{R}^{(0)} = \mathbf{F}_{RF}^{(0)}$, $\mathbf{F}_{BB}^{(0)} = \left(\mathbf{F}_{RF}^{(0)H}\mathbf{F}_{RF}^{(0)}\right)^{-1}\mathbf{F}_{RF}^{(0)H}\mathbf{F}_{opt}$ and $\mathbf{W}^{(0)} = \mathbf{0}$. As for the penalty parameter, it was verified empirically that setting $\rho = 1$ generally achieves good results but it can be fine-tuned for a specific problem setting. Although a maximum number of iterations $Q$ is set for the algorithm, the following stagnation condition can be used for earlier termination

$$\left| f\left(\mathbf{R}^{(t)},\mathbf{F}_{BB}^{(t)}\right) - f\left(\mathbf{R}^{(t+1)},\mathbf{F}_{BB}^{(t+1)}\right) \right| < \tau \tag{19}$$

where $\tau$ is a small positive value. In the simulation results we employed $\tau=0.001$.

### C. Complexity

In the proposed algorithm, the $\mathbf{F}_{RF}^{(t+1)}$ and $\mathbf{F}_{BB}^{(t+1)}$ updates (steps 3 and 4 in table I) are defined using closed-form expressions that encompass several matrix multiplications, sums and an $N_{RF} \times N_{RF}$ matrix inverse (with an assumed complexity order of $\mathcal{O}(N_{RF}^3)$). Both steps require a complexity order of $\mathcal{O}(N_s N_{RF} N_{tx} + N_{RF}^2 N_{tx} + N_{RF}^3)$. The $\mathbf{R}^{(t+1)}$ update (step 5) involves simple elementwise sums and divisions while the dual variable update $\mathbf{W}^{(t+1)}$ (steps 6) comprises only matrix sums. Both steps have complexity orders of $\mathcal{O}(N_{RF} N_{tx})$. Therefore, keeping only the dominant terms and taking into account the number of iterations $Q$, the overall complexity order for the proposed algorithm is $\mathcal{O}(Q(N_s N_{RF} N_{tx} + N_{RF}^2 N_{tx}))$.

TABLE II
COMPLEXITY COMPARISON OF DIFFERENT HYBRID PRECODING METHODS

| Design Method | Number of complex flops per iteration | Total Complexity order |
|---|---|---|
| OMP | $\frac{1}{4}N_{RF}(N_{RF}+1)^2$ <br> $+\frac{1}{6}(N_{RF}+1)(2N_{RF}+1)\left(3N_{tx}-\frac{1}{2}\right)$ <br> $+\frac{1}{2}(N_{RF}+1)\left(4N_{tx}N_s - N_s - \frac{1}{2}\right)$ <br> $+N_{cl}N_{ray}(2N_s N_{tx}+2N_s-1)+3N_{tx}N_s-1$ | $\mathcal{O}\left(N_{cl}N_{ray}N_s N_{RF}N_{tx}+N_{RF}^2 N_{tx}\right)$ |
| HD-AM | $\frac{14}{3}N_{RF}^3 + N_{RF}^2(4N_{tx}-1)$ | $\mathcal{O}\left(QN_{RF}^2 N_{tx}\right)$ |
| HD-LSR | $N_{RF}^3 N_{tx} + N_{RF}^2\left(2N_{tx}N_s + 4N_{tx}^2 - N_{tx}\right) +$ <br> $+N_{RF}\left(N_{tx} - N_{tx}^2 + 2N_{tx}^2 N_s\right)$ | $\mathcal{O}\left(Q\left(N_{RF}^2 N_{tx}^2 + N_{RF}^3 N_{tx}\right)\right)$ |
| Proposed | $2N_{RF}^3 + N_{RF}^2\left(3N_{tx} + 4N_s - \frac{3}{2}\right)$ <br> $+N_{RF}\left(7N_{tx} + 4N_{tx}N_s - 2N_s - \frac{1}{2}\right)$ | $\mathcal{O}\left(Q\left(N_s N_{RF}N_{tx} + N_{RF}^2 N_{tx}\right)\right)$ |

Table II presents the total complexity order of the proposed method and compares it against other existing alternatives namely, OMP based sparse precoder [7] (assuming an angular resolution of $N_{cl}N_{ray}$, i.e., the terminals know the exact angles that make up **H**), hybrid design by alternating minimization (HD-AM) algorithm and hybrid design by least squares relaxation (HD–LSR) algorithm, both from [13]. This table also includes the complexity in terms of complex valued floating-point operations (flops) per iteration for the different methods. As the complexity in flops was not provided in [7] and [13], we evaluated the number of computations required for each individual step of the different algorithms, assuming that each scalar sum, multiplication or division counts as one flop, while an $N_{RF} \times N_{RF}$ matrix inverse counts as $N_{RF}^3$ flops. Note that the complexity of OMP is the average per iteration (a total of $N_{RF}$ iterations are used), and it is the only one that is dictionary based. We also include the computed complexity for two of the scenarios that are evaluated in section IV. It can be seen that in the scenario where $N_s = N_{RF}$, HD-AM has the lowest complexity followed by the proposed approach. For the second scenario, HD-AM cannot be applied, and the proposed approach has clearly the lowest complexity.

## IV. Proposed Hybrid Precoding Algorithm for the Partially Connected Structure

The fully connected structure considered in the previous section requires a phase shifter between each RF chain and each antenna which can incur in substantial implementation complexity. An alternative lower complexity approach is to adopt the partially connected structure from Fig. 2 where each RF chain is connected through phase shifters to a dedicated subset of the antenna array [25][26]. Assuming that $N_{RF}$ chains are used and that each subarray comprises $N_{tx}/N_{RF}$ antennas then, the RF precoder has a block diagonal structure

$$\mathbf{F}_{RF} = \text{blkdiag}\left\{\mathbf{f}^{RF,1},..,\mathbf{f}^{RF,N_{RF}}\right\} = \begin{bmatrix} \mathbf{f}^{RF,1} & 0 & \cdots & 0 \\ 0 & & & \vdots \\ \vdots & & \ddots & 0 \\ 0 & \cdots & 0 & \mathbf{f}^{RF,N_{RF}} \end{bmatrix}, \quad (20)$$

where $\mathbf{f}^{RF,i} = \left[f_1^{RF,i},...,f_{\frac{N_{tx}}{N_{RF}}}^{RF,i}\right]^T \in \mathcal{U}_{N_{tx}/N_{RF},1}$ with $i=1,...,N_{RF}$. The proposed hybrid design approach employed for the fully-connected structure can be directly extended to the partially-connected case. The only modification required concerns the projection in step 5 of the algorithm in Table I which, in this case, should be performed over the set of matrices with the same structure as (20). It is possible, however, to reduce the complexity of the algorithm for this architecture by including the special structure of $\mathbf{F}_{RF}$ directly into the formulation of the problem (3)-(5) resulting

$$\min_{\mathbf{f}^{RF,i},\mathbf{F}_{BB}} \sum_{i=1}^{N_{tx}} \left\| \mathbf{F}_{opt_{i,:}} - f^{RF,\left\lfloor\frac{i-1}{N_{tx}/N_{RF}}\right\rfloor+1}_{\mod\left(i-1,\frac{N_{tx}}{N_{RF}}\right)+1} \mathbf{F}_{BB_{\left\lfloor\frac{i-1}{N_{tx}/N_{RF}}\right\rfloor+1,:}} \right\|_2^2 \quad (21)$$

$$\text{subject to } \mathbf{f}^{RF,i} \in \mathcal{U}_{N_{tx}/N_{RF},1},\ i=1,\ldots,N_{RF} \quad (22)$$

$$\|\mathbf{F}_{BB}\|_F^2 = \frac{N_{RF}}{N_{tx}} N_s. \quad (23)$$

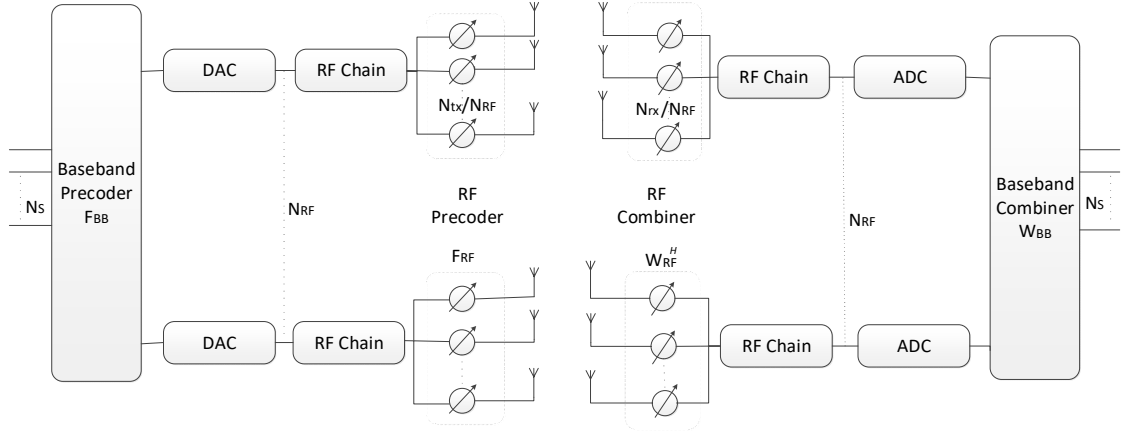

Fig. 2 Hybrid analog/digital precoder and combiner with partially-connected structure.

Applying a similar derivation to the one used in section III.A results in the following expressions for steps 3 to 6 of the algorithm in Table I

$$\left(f_j^{RF,i}\right)^{(t+1)} = \frac{\mathbf{F}_{opt_{(i-1)\frac{N_{tx}}{N_{RF}}+j,:}} \mathbf{F}_{BB_{i,:}}^{(t)\,H} + \rho\left(r_j^{i\,(t)} - w_j^{i\,(t)}\right)}{\left\|\mathbf{F}_{BB_{i,:}}^{(t)}\right\|^2 + \rho},\ i=1,\ldots,N_{RF}, j=1,\ldots,\frac{N_{tx}}{N_{RF}} \quad (24)$$

$$\mathbf{F}_{BB_{i,:}}^{(t+1)} = \left\|\mathbf{f}^{RF,i\,(t+1)}\right\|^{-2} \left(\mathbf{f}^{RF,i\,(t+1)}\right)^H \mathbf{F}_{opt_{(i-1)\frac{N_{tx}}{N_{RF}}+1:i\frac{N_{tx}}{N_{RF}},:}},\ i=1,\ldots,N_{RF} \quad (25)$$

$$\mathbf{r}^{i\,(t+1)} = \left(\mathbf{f}^{RF,i\,(t+1)} + \mathbf{w}^{i\,(t)}\right) \oslash \left|\mathbf{f}^{RF,i\,(t+1)} + \mathbf{w}^{i\,(t)}\right|,\ i=1,\ldots,N_{RF} \quad (26)$$

$$\mathbf{w}^{i\,(t+1)} = \mathbf{w}^{i\,(t)} + \mathbf{f}^{RF,i\,(t+1)} - \mathbf{r}^{i\,(t+1)},\ i=1,\ldots,N_{RF}. \quad (27)$$

TABLE III

ITERATIVE HYBRID DESIGN ALGORITHM FOR THE PARTIALLY-CONNECTED STRUCTURE

1: **Input:** $\mathbf{F}_{opt}, \mathbf{f}^{RF,i^{(0)}}, \mathbf{F}_{BB}^{(0)}, \mathbf{r}^{i^{(0)}}, \mathbf{w}^{i^{(0)}}, \rho, Q$
2: **for** $t=0,1,\ldots Q-1$ **do**
3:    Compute $\mathbf{f}^{RF,i^{(t+1)}}$ using (24) for all $i=1,\ldots,N_{RF}$.
4:    Compute $\mathbf{F}_{BB}^{(t+1)}$ using (25).
5:    Compute $\mathbf{r}^{i^{(t+1)}}$ via projection (26) for all $i=1,\ldots,N_{RF}$.
6:    Update $\mathbf{w}^{i^{(t+1)}}$ using (27) for all $i=1,\ldots,N_{RF}$.
7: **end for**.
8: $\hat{\mathbf{F}}_{RF} \leftarrow \text{blkdiag}\left\{\mathbf{r}^{1^{(Q)}},..,\mathbf{r}^{N_{RF}^{(Q)}}\right\}$
9: $\hat{\mathbf{F}}_{BB} \leftarrow \text{blkdiag}\left\{\left\|\mathbf{r}^{1^{(Q)}}\right\|^{-2},..,\left\|\mathbf{r}^{N_{RF}^{(Q)}}\right\|^{-2}\right\}\hat{\mathbf{F}}_{RF}^{H}\mathbf{F}_{opt}$
10: $\hat{\mathbf{F}}_{BB} \leftarrow \sqrt{\dfrac{N_{RF}}{N_{tx}}N_s}\left\|\hat{\mathbf{F}}_{BB}\right\|_F^{-1}\hat{\mathbf{F}}_{BB}$ (for precoder)
11: **Output:** $\hat{\mathbf{F}}_{BB}, \hat{\mathbf{F}}_{RF}$.

Note that instead of working with matrices for the auxiliary variable **R** and dual variable **W**, this derivation uses vectors of length $N_{tx}/N_{RF}$, namely $\mathbf{r}^i$ and $\mathbf{w}^i$ with $i=1,\ldots,N_{RF}$. The simplified algorithm is summarized in table III.

## V. Hybrid Precoding Algorithm for OFDM-based mmWave systems

The system model adopted in the previous sections assumed a narrowband mmWave channel. However, due to the large bandwidth available in mmWave bands, it is likely that practical MIMO systems will have to operate in scenarios where the channel is frequency selective. Therefore, to cope with the resulting multipath fading, multicarrier schemes such as OFDM are often employed [23][27]. In this section we extend the proposed hybrid precoder/combiner design to a OFDM-based mmWave system as shown in Fig. 3. Assuming the system is working with blocks of $K$ subcarriers in a wideband mmWave channel where cluster $i$ has a time delay of $i$-1, as in [23][27], then the frequency domain channel matrix at

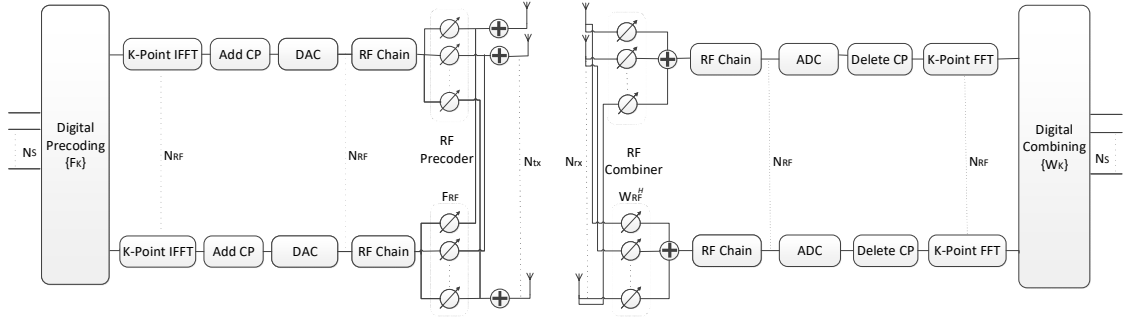

Fig. 3 Hybrid analog/digital precoder and combiner for a OFDM-based mmWave system.

the $k^{\text{th}}$ subcarrier can be written as

$$\mathbf{H}[k] = \gamma \sum_{i=1}^{N_{cl}} \sum_{l=1}^{N_{ray}} \alpha_{i,l} \mathbf{a}_r\left(\phi_{i,l}^r, \theta_{i,l}^r\right) \mathbf{a}_t\left(\phi_{i,l}^t, \theta_{i,l}^t\right)^H e^{-j2\pi(i-1)k/K} . \quad (28)$$

As shown in Fig. 3, we consider that digital baseband precoding is applied before the IFFT, i.e., at the frequency domain for each subcarrier, while a common analog beamforming is applied in the time domain to the whole block. At the receiver, a similar approach is used, with the combiner applied after the FFT. Therefore, the received signal at subcarrier $k$ after the combiner can be expressed as

$$\mathbf{y}[k] = \sqrt{\varepsilon} \mathbf{W}_{\text{BB}}^H[k] \mathbf{W}_{\text{RF}}^H \mathbf{H}[k] \mathbf{F}_{\text{RF}} \mathbf{F}_{\text{BB}}[k] \mathbf{s}[k] + \mathbf{W}_{\text{BB}}^H[k] \mathbf{W}_{\text{RF}}^H \mathbf{n}[k] . \quad (29)$$

In this case, the objective is to find $K$ different digital precoding and combining matrices, $\mathbf{F}_{\text{BB}}[k]$ and $\mathbf{W}_{\text{BB}}[k]$, one common RF precoding, $\mathbf{F}_{\text{RF}}$, and one common combining matrix, $\mathbf{W}_{\text{RF}}$. The hybrid design problem can be formulated as an extension of the matrix factorization problem (3)-(5) used for the narrowband case [23][27] and applied separately to the precoder and combiner sides (in the later no total power constraint is required), namely

$$\min_{\mathbf{F}_{\text{RF}}, \mathbf{F}_{\text{BB}}[k]} \sum_{k=0}^{K-1} \left\| \mathbf{F}_{opt}[k] - \mathbf{F}_{\text{RF}} \mathbf{F}_{\text{BB}}[k] \right\|_F^2 \quad (30)$$

$$\text{subject to } \mathbf{F}_{\text{RF}} \in \mathcal{U}_{N_{tx}, N_{RF}} \quad (31)$$

$$\left\| \mathbf{F}_{\text{RF}} \mathbf{F}_{\text{BB}}[k] \right\|_F^2 = N_s, \ k = 0, \ldots, K-1. \quad (32)$$

This formulation allows us to apply ADMM using the same approach of section III.A. An algorithm identical to the one in table I is obtained with the only differences lying in steps 3 and 4 which correspond to the minimization over $\mathbf{F}_{RF}$ and over $\mathbf{F}_{BB}[k]$. In this case, it is simple to show that these two minimization steps are accomplished using the following expressions:

$$\mathbf{F}_{RF}^{(t+1)} = \left[ \sum_{k=0}^{K-1} \mathbf{F}_{opt}[k]\mathbf{F}_{BB}^{(t)}[k]^H + \rho\left(\mathbf{R}^{(t)} - \mathbf{W}^{(t)}\right) \right] \left( \sum_{k=0}^{K-1} \mathbf{F}_{BB}^{(t)}[k]\mathbf{F}_{BB}^{(t)}[k]^H + \rho\mathbf{I}_{N_{RF}} \right)^{-1} \quad (33)$$

$$\mathbf{F}_{BB}[k]^{(t+1)} = \left(\mathbf{F}_{RF}^{(t+1)H}\mathbf{F}_{RF}^{(t+1)}\right)^{-1}\mathbf{F}_{RF}^{(t+1)H}\mathbf{F}_{opt}[k], \; k=0,\ldots,K-1. \quad (34)$$

It is important to highlight that the proposed hybrid design algorithm for mmWave MIMO-OFDM systems is the direct extension of the narrowband version and, in fact, reduces to the latter when $K=1$.

## VI. Numerical Results

In this section we evaluate the performance of the proposed algorithm. We consider channel model (6) with $\alpha_{i,l} \sim \mathcal{CN}(0,1)$ and Gaussian distributed angles of departure and arrival. The means of the azimuth and elevation angles for each cluster are uniformly random distributed in $[0, 2\pi]$ and the angular spreads are all constant and equal with values of $\sigma_\phi^t = \sigma_\phi^r = \sigma_\theta^t = \sigma_\theta^r = 10^o$. A total of $N_{cl}=8$ scattering clusters with $N_{ray}=10$ paths each are assumed. In the simulations we consider a uniform square planar array with $\sqrt{N_{tx}} \times \sqrt{N_{tx}}$ antenna elements at the transmitter and $\sqrt{N_{rx}} \times \sqrt{N_{rx}}$ at the receiver. The respective array response vectors are given by

$$\mathbf{a}_{t,r}\left(\phi_{i,l}^{t/r},\theta_{i,l}^{t/r}\right) = \frac{1}{\sqrt{N_{tx/rx}}}\left[1,\ldots,e^{j\frac{2\pi}{\lambda}d\left(p\sin\phi_{i,l}^{t/r}\sin\theta_{i,l}^{t/r}+q\cos\theta_{i,l}^{t/r}\right)},$$

$$...,e^{j\frac{2\pi}{\lambda}d\left(\left(\sqrt{N_{tx/rx}}-1\right)\sin\phi_{i,l}^{t/r}\sin\theta_{i,l}^{t/r}+\left(\sqrt{N_{tx/rx}}-1\right)\cos\theta_{i,l}^{t/r}\right)}\Bigg]^T, \qquad (35)$$

where $p,q = 0,...,\sqrt{N_{tx/rx}}-1$ are the antenna indices, $\lambda$ is the signal wavelength and $d$ is the inter-element spacing (assumed to be $d = \lambda/2$). Each point was computed with 5000 independent Monte Carlo runs.

### A. Hybrid design for narrowband mmWave channels

First, we evaluate the performance of the proposed algorithms in the case of a narrowband mmWave system. Fig. 4 shows the spectral efficiency versus the signal to noise ratio (SNR), defined as $SNR = \varepsilon/\sigma_n^2$, achieved by the proposed algorithm ($Q$=30), OMP based sparse method [7], HD-AM (10 iterations) and HD–LSR (20 iterations) (both from [13]). Note that these three methods were selected as references since they are well-known methods capable of near optimal performance (specially HD-AM and HD-LSR). It is assumed that, given the unconstrained data rate maximizing precoder/combiner matrices, each algorithm is applied for both precoding and combiner design. Even though we are focusing on point-to-point MIMO transmissions, it is interesting to compare the behaviour of a precoder/combiner designed specifically for multiuser scenarios. Therefore we also included a curve of the hybrid minimum mean-squared error (MMSE) algorithm from [18]. This algorithm has a complexity order of $\mathcal{O}\left(N_{tx}^3 N_{RF}\right)$ which compares against $\mathcal{O}\left(Q\left(N_s N_{RF} N_{tx} + N_{RF}^2 N_{tx}\right)\right)$ (with $Q$=30 in these results) for the algorithm proposed in this paper,. When the number of streams is the same as the number of RF chains, $N_s = N_{RF}$, all methods achieve spectral efficiencies close to the optimal fully digital, apart from the sparse based approach and the hybrid MMSE algorithm. When $N_s < N_{RF}$ the proposed method and HD-LSR achieve the best performances but the latter has a much higher total computational cost since it requires 350581760 complex valued operations while the proposed method needs 3703080 (the sparse

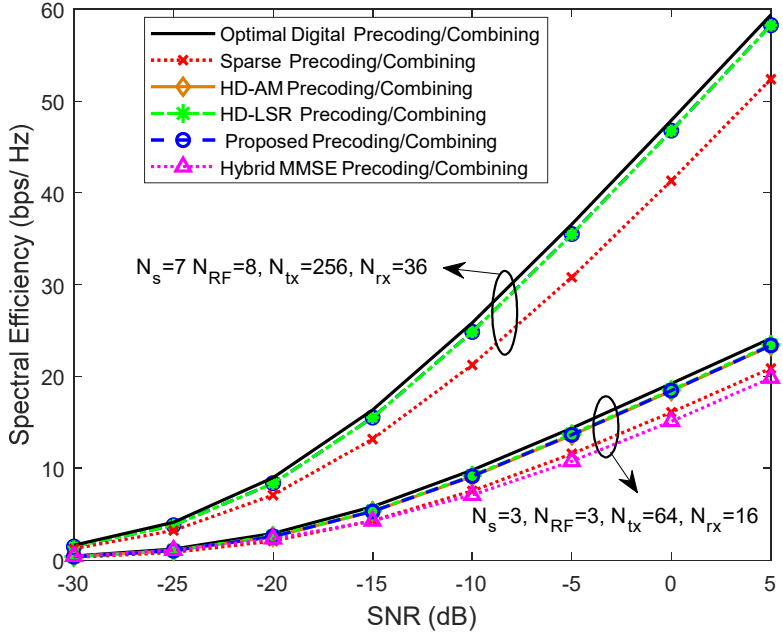

Fig. 4 Spectral efficiency versus SNR achieved by different methods for a narrowband mmWave system.

TABLE IV

AVERAGE COMPUTATION TIMES (MILISECONDS) PER PAIR OF PRECODERS, $\mathbf{F}_{RF}$, $\mathbf{F}_{BB}$, AND COMBINERS, $\mathbf{W}_{RF}$, $\mathbf{W}_{BB}$, FOR THE DIFFERENT ALGORITHMS OF FIG. 4

| Scenario | Sparse | HD-AM | HD-LSR | Hybrid MMSE | Proposed |
|---|---|---|---|---|---|
| $N_S$=7, $N_{RF}$=8, $N_{tx}$=256, $N_{rx}$=36 | 6.4 | - | 1753 | - | 33 |
| $N_S$=3, $N_{RF}$=3, $N_{tx}$=64, $N_{rx}$=16) | 1.4 | 1 | 82 | 1.4 | 2.7 |

precoder needs 2760720). Note that no HD-AM curve is included for $N_s < N_{RF}$ as it can only be applied when $N_s = N_{RF}$. As an additional comparison of the complexities of the different methods, table IV presents the average computation time required for calculating the precoder and combiner matrices, i.e., $\mathbf{F}_{RF}$, $\mathbf{F}_{BB}$, $\mathbf{W}_{RF}$ and $\mathbf{W}_{BB}$. These times include all the steps, iterations and initializations applied in each method (including initial optimal digital matrix computation). The simulations were run using Matlab Release 2017b on a 3.70 GHz Intel i7 machine with 12 threads. While the sparse based method and the hybrid MMSE are faster than the proposed algorithm, their performances are worse. As for HD-AM, it is also faster and achieves the same performance. However, as previously stated, it is restricted to scenarios where $N_s = N_{RF}$ whereas the proposed approach can be applied to any configuration of antennas, RF chains and streams.

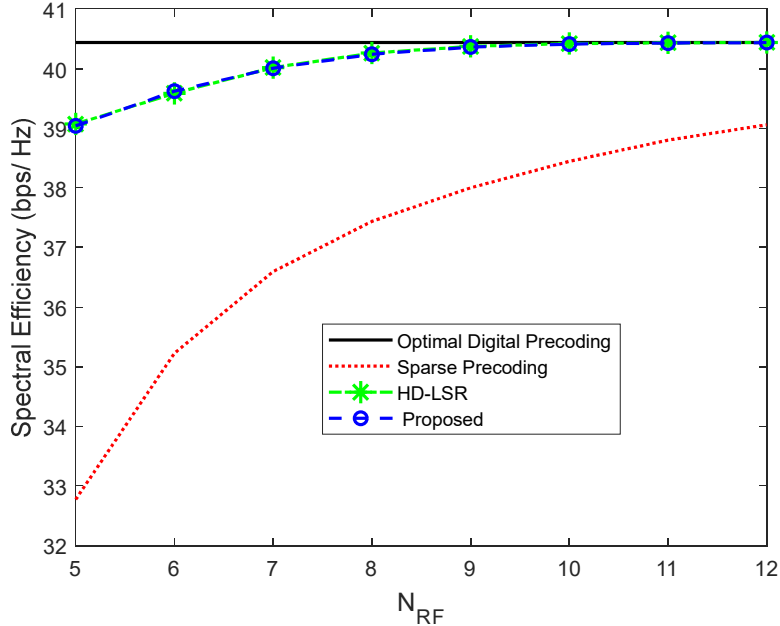

Fig. 5. Spectral efficiency versus number of RF chains for $N_s$=5, $N_{tx}$=256, $N_{rx}$=36 and *SNR*=0 dB.

Fig. 5 shows the spectral efficiency versus $N_{RF}$ for $N_s = 5$, $N_{tx}$=256, $N_{rx}$=36 and *SNR*=0 dB. Increasing $N_{RF}$ reduces the gap to the optimal fully digital design for all methods, with the proposed one and HD-LSR achieving the highest spectral efficiencies and basically matching the performance of the optimal design when $N_{RF} \geq 10$ (with HD-LSR always having a much higher complexity cost).

In Fig. 6 we present the spectral efficiencies for a scenario with $N_s = N_{RF} = 2$, $N_{tx}$=100, $N_{rx}$=16 where we also include results for the partially-connected structure. Regarding this architecture, the SIC-based approach presented in [25] is used as a benchmark. It can be seen that the fully-connected architectures using HD-AM, HD-LSR and the proposed algorithm show a similar behaviour as in the previous results, performing close to the optimal digital design. As for the partially-connected architectures, the SIC-based method is able to achieve similar performance to the fully-connected sparse-based design and the hybrid MMSE, while the proposed approach is even able to outperform those.

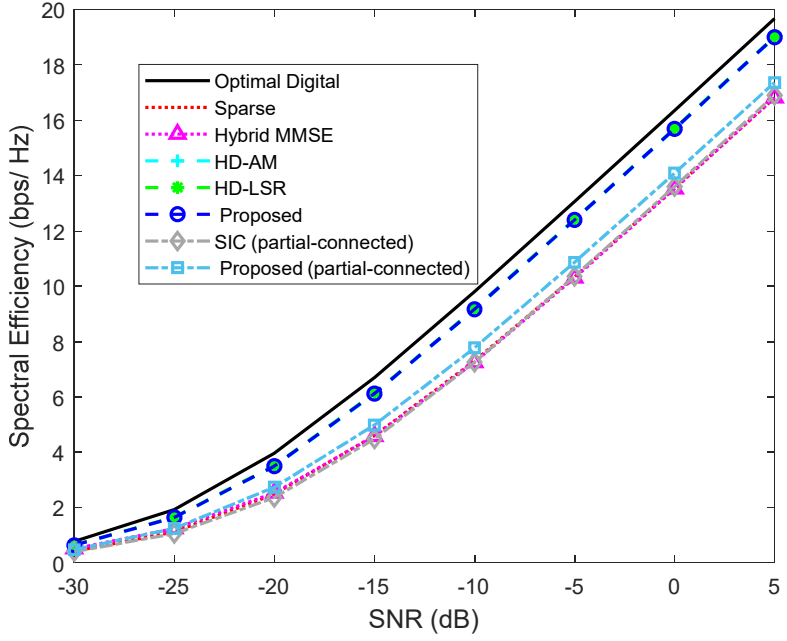

Fig. 6 Spectral efficiency versus SNR achieved by different methods for a narrowband mmWave system with $N_s = 2$, $N_{RF} = 2$, $N_{tx} = 100$ and $N_{rx} = 16$.

### B. Hybrid design for OFDM-based mmWave systems

In this subsection we evaluate the performance of the proposed hybrid design algorithm on a OFDM-based mmWave system operating in a frequency selective channel. We assume that the number of subcarriers is $K=128$, $N_s = 3$, $N_{RF} = 4$, $N_{tx} = 144$ and $N_{rx} = 36$. The number of clusters and rays per cluster are the same as in the previous scenarios. Fig. 7 shows the spectral efficiency versus SNR of the proposed approach and compares it against the PE-AltMin algorithm from [27] and the OMP-based hybrid algorithm from [23]. Even though the gap to the full digital precoder/combiner is wider than in the narrowband case (as expected since in the hybrid design the RF precoder/combiner is the same for all subcarriers),

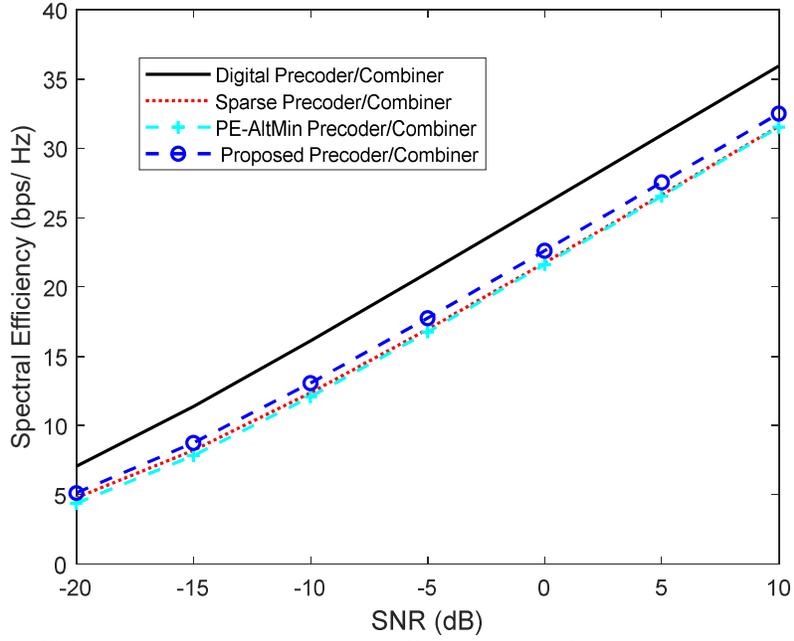

Fig. 7 Spectral efficiency versus SNR achieved by different methods for a mmWave MIMO-OFDM system with $K$=128, $N_s = 3$, $N_{RF}$ =4, $N_{tx}$ =144 and $N_{rx}$ =36.

the proposed approach manages to achieve a better performance than the other hybrid algorithms.

## VII. Conclusions

In this work we have addressed the hybrid design problem for spatial multiplexing in mmWave MIMO systems using an augmented Lagrangian based decomposition method. The adopted approach results in an iterative algorithm comprising a sequence of smaller subproblems with straightforward solutions. The precoding/combining design algorithm can work with a broad range of configuration of antennas, RF chains and data streams. Furthermore, different versions of the algorithm were proposed for the fully-connected and partially-connected hybrid precoding structures, as well as for a OFDM-based mmWave system operating in frequency selective channels. Simulation results show that the proposed method is capable of performing close to the optimal fully digital design, with a better performance-complexity trade-off than other existing methods, in particular when the number of streams and RF chains are different. In the future it will be interesting to extend the approach to other hybrid design settings, like multiuser uplink/downlink and multi-cell, and also consider the impact of channel estimation and feedback.

## VIII. Acknowledgments

This work is funded by FCT/MEC through national funds and when applicable co-funded by FEDER – PT2020 partnership agreement under the project UID/EEA/50008/2019 and TUBITAK/0002/2014.